\documentclass[usegraphicx,usenatbib,letterpaper]{emulateapj}

\usepackage{ulem}
\usepackage{color}
\usepackage{graphics,graphicx}
\usepackage{times} 
\usepackage{amssymb}
\usepackage{amsmath}
\usepackage{natbib}
\usepackage{url}
\usepackage{multirow}
\usepackage{ctable}
\usepackage{threeparttable}
\newif\ifAMStwofonts
\AMStwofontstrue

\def\chandra{{\it Chandra}}

\def\epicmos1{{EPIC-MOS1}}
\def\epicmos2{{EPIC-MOS2}}
\def\epicmos{{EPIC-MOS}}

\def\nustar{{\it NuSTAR}}

\def\deg{$^{\circ}$}

\def\H0{{\rm ~km~s^{-1}~Mpc^{-1}}}

\def\kev{\hbox{\rm keV}}

\def\chisq{{$\chi^{2}$}}

\def\xspec{\hbox{\small XSPEC}}

\def\heasoft{\hbox{\rm{\small HEASOFT}}}

\def\flx2xsp{\rm{\small FLX2XSP}}

\def\grid25{\hbox{\rm{\small GRID25}}}

\def\relline{\rm{\small RELLINE}}

\def\ka{$\rm{K}\alpha$}

\def\eg{{\it e.g.}}

\def\la{\mathrel{\hbox{\rlap{\hbox{\lower4pt\hbox{$\sim$}}}{\raise2pt\hbox{$<$}}}}}
\def\ga{\mathrel{\hbox{\rlap{\hbox{\lower4pt\hbox{$\sim$}}}{\raise2pt\hbox{$>$}}}}}

\def\d25{D$_{25}$}

\def\.25{0.25 keV\thinspace}

\def\rg{$R_{\rm G}$}

\def\nsims{\rm{10,000}}
\def\ngtr{\rm{7}}

\shorttitle{Broad Iron Emission from Gravitationally Lensed Quasars}
\shortauthors{D.~J. Walton et al.}

\begin{document}

\title{Broad Iron Emission from Gravitationally Lensed Quasars Observed by \textit{Chandra}}

\author{D. J. Walton\altaffilmark{1,2},
M. T. Reynolds\altaffilmark{3},
J. M. Miller\altaffilmark{3},
R. C. Reis\altaffilmark{3},
D. Stern\altaffilmark{1},
F. A. Harrison\altaffilmark{2}
}
\affil{
$^{1}$ Jet Propulsion Laboratory, California Institute of Technology, Pasadena, CA 91109, USA \\
$^{2}$ Space Radiation Laboratory, California Institute of Technology, Pasadena, CA 91125, USA \\
$^{3}$ Department of Astronomy, University of Michigan, 1085 S. University Ave., Ann Arbor, MI, 49109-1107, USA \\
}

\begin{abstract}
Recent work has demonstrated the potential of gravitationally lensed quasars
to extend measurements of black hole spin out to high-redshift with the current
generation of X-ray observatories. Here we present an analysis of a large sample
of 27 lensed quasars in the redshift range $1.0\lesssim{z}\lesssim4.5$ observed
with \chandra, utilizing over 1.6\,Ms of total observing time, focusing on the
rest-frame iron K emission from these sources. Although the X-ray
signal-to-noise (S/N) currently available does not permit the detection of iron 
emission from the inner accretion disk in individual cases in our sample,
we find significant structure in the stacked residuals. In addition to the narrow
core, seen almost ubiquitously in local AGN, we find evidence for an additional
underlying broad  component from the inner accretion disk, with a clear red wing
to the emission profile. Based on simulations, we find the detection of this broader
component to be significant at greater than the $3\sigma$ level. This implies that
iron emission from the inner disk is relatively common in the population of lensed
quasars, and in turn further demonstrates that, with additional observations, this
population represents an opportunity to significantly extend the sample of AGN
spin measurements out to high-redshift.
\end{abstract}


\section{Introduction}

Information regarding the manner in which the supermassive black holes (SMBHs)
powering active galactic nuclei (AGN) grew is encoded in the distribution of their
angular momenta $J$ (or more specifically their `spin',
$a^{*}\equiv{J}c/GM_{\rm{BH}}^{2}$; \citealt{King06, Berti08, Volonteri13}). For
example, if these black holes grew through prolonged episodes of coherent
accretion, we should see a preference for rapidly rotating  black holes. In contrast,
if they grew through a series of chaotic mergers and  accretion events we should
instead see a lower value for the average black hole spin.

AGN spin measurements are anchored in X-ray spectroscopy, and rely primarily
on measuring the relativistic distortions imprinted on fluorescent line emission
from the inner disk (\citealt{Fabian89, kdblur}), which result in intrinsically narrow
emission lines being broadened and skewed into a characteristic `diskline' profile.
Although lines from a variety of elements are naturally produced in the resulting
`reflected' emission when the optically thick accretion disk is irradiated by
high-energy X-rays, the most prominent is the iron \ka\ line at $\sim$6--7
\kev\ (depending on ionisation state), owing to its high cosmic abundance and
fluorescent yield (\citealt{George91}). It is also best suited for measuring these
relativistic distortions, being relatively isolated from other strong emission lines.
For recent reviews on relativistic disk reflection, see \cite{Miller07rev} and
\cite{Reynolds13rev}.

Spin estimates for a growing sample of $\sim$20--30 local ($z\sim0$) AGN
have been obtained through study of these features, (\eg\ \citealt{Walton13spin,
Fabian13iras, Risaliti13nat}), and already suggest that many black holes have high
spin, although the sample is not yet well defined in a statistical sense
(\citealt{Brenneman13book, Reynolds13rev}). As models linking the cosmic
growth of SMBHs and galaxy formation become more advanced (\citealt{Dubois14,
Sesana14}), detailed comparison with observation will require knowledge of the
spin distribution as a function of redshift, extending out to and beyond the peak
of AGN activity ($z\sim2$; \citealt{Richards06}). However, current X-ray
instrumentation does not have the sensitivity required to undertake such
measurements for typical AGN at these redshifts.

Recently, however, we have demonstrated that strongly lensed quasars offer a
rare opportunity to obtain spin measurements from objects at cosmologically
interesting redshifts with current instrumentation, owing to the combination of
the multiple images observed and the amplification of the intrinsic emission by
the lens. The two cases with sufficient signal-to-noise (S/N), RXJ1131-1231
($z=0.658$) and Q2237+0305 (aka the Einstein Cross; $z=1.695$), exhibit
relativistic iron disklines similar to those observed in local Seyfert galaxies,
allowing us to infer that both sources host rapidly rotating black holes
(\citealt{Reis14nat, Reynolds14}). Here, we examine the iron emission from a
sample of lensed quasars with X-ray observations that currently have lower S/N,
to investigate whether such iron disklines are common among this population,
and whether it could potentially facilitate further high-redshift spin
measurements.

\section{Sample Selection and Data Reduction}
\label{sec_red}

We selected a sample of strong gravitationally lensed quasars from the publicly
available CASTLES database\footnote{\url{http://www.cfa.harvard.edu/castles/}}
(numbering 100 lensed quasars in total as of August 2014). We select all those
systems characterized as grade A (i.e., confirmed multiply imaged strong
gravitationally lensed quasars\footnote{per the CASTLES website: ``I'd bet
\textit{my} life on it''.}) with redshift determinations for both the lens and the
quasar (\eg, there are eight grade A systems with no secure redshift for the
lens, which we exclude from our analysis). Additionally, both RXJ1131-1231
(\citealt{Reis14nat}) and Q2237+0305 (\citealt{Reynolds14}) are excluded from
the sample, having individually been shown to exhibit prominent
relativistic lines (see also \citealt{Dai03, Chen12}). We then searched for sources
with imaging observations obtained with the \chandra\ observatory
(\citealt{CHANDRA}), restricting ourselves to observations within $1'$ of the S3
aimpoint on the ACIS detector (\citealt{CHANDRA_ACIS}). This resulted in an
initial sample of 35 systems. The majority of this sample are lensed by
foreground galaxies, but three sources are lensed by galaxy clusters (Q0957+561,
SDSSJ1004+4112 and SDSSJ1029+2623; see \citealt{Chartas02}, \citealt{Ota06},
and \citealt{Ota12}, respectively), and two are lensed by galaxies within faint X-ray
emitting galaxy groups (PG1115+080 and B1422+231; \citealt{Grant04}). We note
however that previous work has shown that the cluster/group emission peaks at
$\lesssim$1\,\kev\ in the observed frame, and that it makes a negligible
contribution to the iron bandpass for the quasars in these cases.

All observations were re-processed in \textsc{ciao
v4.5}\footnote{\url{http://cxc.harvard.edu/ciao}}, with the \textsc{EDSER}
algorithm enabled and with the latest \textit{Chandra} calibration files. The
resulting event files were re-binned to 1/8$\rm^{th}$ of the native ACIS pixel
size before smoothing with a Gaussian of 0.25$\arcsec$ (FWHM). In those cases
where individual images were resolved, spectra were extracted from
0.5$\arcsec$ radius regions centered on each individual sub-pixel image via the
\textsc{specextract} script with PSF correction enabled. If this was not possible
(\eg\ due to the quasar images being too close for even \chandra\ to
resolve, or low count rates precluding centroiding), larger extraction regions
encompassing multiple images were utilized. All spectra for each source were
subsequently combined using the {\small{COMBINE\_SPECTRA}} script before
grouping to S/N=3 per spectral bin with {\small{DMGROUP}}; we verified that we
obtain consistent results using an initial binning of S/N=4. The spectra and
background files for all sources were then exported to
\xspec\footnote{\url{http://heasarc.gsfc.nasa.gov/xanadu/xspec/}} for spectral
analysis (v12.8.0m; \citealt{xspec}). In the observed frame, the \chandra\ data
are modeled over the 0.35--8.0\,\kev\ bandpass.

\section{Stacking Analysis}
\label{sec_stack}

To investigate the iron K emission from our sample we stacked the
2--10\,\kev\ data to produce a single, average spectrum, owing to the low
photon statistics currently available for the individual sources in our sample (see
Table \ref{tab_sample}), adopting an approach similar to the residuals-based
methodology outlined in \cite{Chaudhary12}, as described below (cf.
\citealt{Nandra97, Guainazzi06}). We limit ourselves to this method, rather than
additionally redshift-correcting and stacking the spectra and instrumental
responses directly, as \cite{Chaudhary12} find in their analysis of 2XMM sources
that the two methods return consistent results. Furthermore, this method has
the advantage that the extinction towards each source can be treated individually.

\begin{table}
  \caption{The sample of 27 gravitationally lensed quasars observed by Chandra
  included in our analysis.}
\begin{center}
\begin{tabular}{l c c c c c}
\hline
\hline
\\[-0.25cm]
Source & $z_{\rm{qso}}$ & $z_{\rm{lens}}$ & $N_{\rm{i}}$\tmark[a] & Total & 2-10\,\kev\ \\
& & & & exposure\tmark[b] & counts\tmark[c] \\
\\[-0.3cm]
\hline
\hline
\\[-0.1cm]
B 1152+199 & 1.019 & 0.439 & 2 & 27 & 1927 \\
\\[-0.225cm]
SDSS J1226-0006 & 1.12 & 0.52 & 2 & 5 & 48 \\
\\[-0.225cm]
FBQ 0951+2535 & 1.24 & 0.26 & 2 & 35 & 132 \\
\\[-0.225cm]
Q 0158-4325 & 1.29 & 0.317 & 2 & 35 & 550 \\
\\[-0.225cm]
B 0712+472 & 1.34 & 0.41 & 4 & 99 & 627 \\
\\[-0.225cm]
SBS 0909+532 & 1.377 & 0.83 & 2 & 20 & 1520 \\
\\[-0.225cm]
Q 0957+561\tmark[d] & 1.41 & 0.36 & 2 & 37 & 10919 \\
\\[-0.225cm]
SDSS J0924+0219 & 1.524 & 0.39 & 2 & 128 & 735 \\
\\[-0.225cm]
B 1600+434 & 1.59 & 0.41 & 2 & 31 & 170 \\
\\[-0.225cm]
HE 0047-1756 & 1.66 & 0.41 & 2 & 40 & 1017 \\
\\[-0.225cm]
WFI J2033-4723 & 1.66 & 0.661 & 4 & 30 & 378 \\
\\[-0.225cm]
HE 0435-1223 & 1.689 & 0.46 & 4 & 110 & 1699 \\
\\[-0.225cm]
PG 1115+080 & 1.72 & 0.31 & 4 & 156 & 6353 \\
\\[-0.225cm]
SDSS J1004+4112\tmark[d] & 1.734 & 0.68 & 4/5\tmark[e] & 155 & 7745 \\
\\[-0.225cm]
HE 0230-2130 & 2.162 & 0.52 & 4 & 45 & 1063 \\
\\[-0.225cm]
SDSS J1029+2623\tmark[d] & 2.197 & 0.55 & 2 & 57 & 1715 \\
\\[-0.225cm]
PMN J0134-0931 & 2.216 & 0.77 & 4/5\tmark[e] & 1 & 73 \\
\\[-0.225cm]
HE 1104-1805 & 2.32 & 0.73 & 2 & 131 & 2192 \\
\\[-0.225cm]
SDSS J1138+0314 & 2.438 & 0.445 & 4 & 51 & 160 \\
\\[-0.225cm]
MG J0414+0534 & 2.639 & 0.9854 & 4 & 200 & 8696 \\
\\[-0.225cm]
QSO J1004+1229 & 2.65 & 0.95 & 2 & 19 & 36 \\
\\[-0.225cm]
Q 0142-100 & 2.72 & 0.49 & 2 & 14 & 407 \\
\\[-0.225cm]
LBQS 1009-0252 & 2.74 & 0.87 & 2 & 10 & 94 \\
\\[-0.225cm]
RX J0911.4+0551 & 2.80 & 0.77 & 4 & 65 & 347 \\
\\[-0.225cm]
HS 0818+1227 & 3.115 & 0.39 & 2 & 20 & 197 \\
\\[-0.225cm]
B 1422+231 & 3.62 & 0.34 & 4 & 125 & 9509 \\
\\[-0.225cm]
BRI 0952-0115 & 4.5 & 0.632 & 2 & 20 & 42 \\
\\[-0.2cm]
\hline
\\[-0.2cm]
Total & & & & 1666 & 58351 \\
\hline
\hline
\\[-0.2cm]
\multicolumn{6}{l}{$^{a}$ The number of lensed images of each quasar included in
our sample} \\
\multicolumn{6}{l}{$^{b}$ Total \chandra\ exposure in ks} \\
\multicolumn{6}{l}{$^{c}$ Assessed in the quasar rest-frame} \\
\multicolumn{6}{l}{$^{d}$ Lensed by a cluster of galaxies} \\
\multicolumn{6}{l}{$^{e}$ Although there are five lensed images in the optical, only
four have} \\
\multicolumn{6}{l}{robust X-ray detections.} \\
\end{tabular}
\label{tab_sample}
\end{center}
\end{table}

For each individual source we fit a simple absorbed powerlaw continuum to the
full observed bandpass, excluding the rest-frame 3.5--7.5\,\kev\ energy range
where the iron emission may contribute, and then determined the data/model
ratio for the 2--10\,\kev\ rest-frame bandpass. This model includes both a
fixed Galactic absorption component (\citealt{NH}) and a second neutral
absorption component at the redshift of the source that was free to vary, and
we limited the photon index to $1.3\leq\Gamma\leq3.0$. Of the 35 sources in
our initial sample, 8 did not have sufficient data to fit the continuum above and
below the iron bandpass. These were subsequently excluded from our analysis,
resulting in a final sample of 27 lensed QSOs; the basic details of this sample
are presented in Table \ref{tab_sample}.

We then corrected the ratio spectra obtained for the source redshift, and
re-sampled to a common rest-frame energy binning. When re-sampling the
ratio spectra, we assumed the counts included in each of the original bins to be
evenly distributed across the energy range covered by that bin, and
present the results with this analysis. However, identical results are obtained if
we instead assume a powerlaw distribution, adopting the photon index obtained
from our initial continuum fits for each source respectively. To determine the
data/model ratio for each of the new bins, the ratios in each of the overlapping
original bins were weighted by the fraction  of the energy range of the new bin
that they contribute, and then averaged. The fractional uncertainties were
determined from the counts in each new bin, assuming Poisson statistics. We
re-sampled the 2--10\,\kev\ bandpass into 50 evenly spaced energy bins of
width 160\,eV, slightly broader than the ACIS spectral resolution at the
observed iron energies, given the redshift range probed. Finally, we stacked the
re-sampled  ratio spectra by calculating the weighted average of the individual
ratios in each of the new bins. The total number of counts contributing to each
of our re-sampled energy bins is always greater than 200. In order to test our
re-sampling/stacking analysis, we simulated a series of spectra at different
redshifts with a common model, consisting of a powerlaw continuum and a
diskline emission line profile, applied our analysis to these simulated spectra,
and verified that the stacked residuals reproduced the input line profile. We also
confirmed that the results presented below do not depend on our re-sampling,
again obtaining consistent results with an analysis re-sampling to 40 bins
(200\,eV width) instead.

\section{Iron Emission}
\label{sec_fek}

\begin{figure}
\hspace*{-0.6cm}
\epsscale{1.15}
\plotone{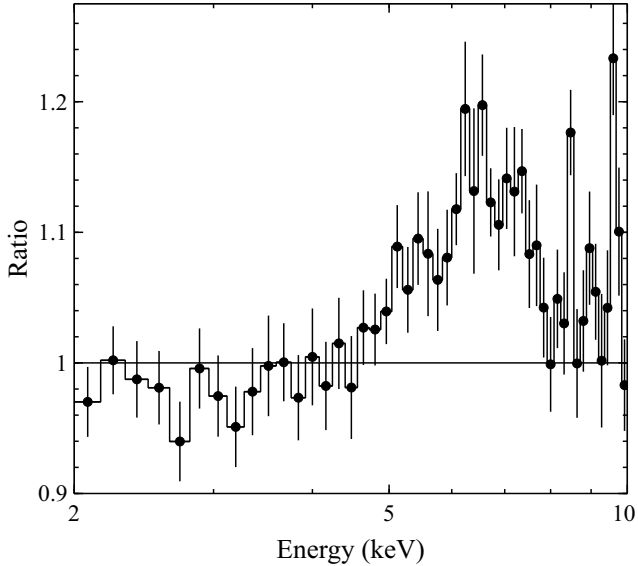}
\caption{
The stacked data/model residuals in the Fe K bandpass for the sample of 27
lensed quasars observed by \chandra, modeled outside the Fe K bandpass with a
simple absorbed powerlaw continuum. The stacked iron emission profile exhibits
a combination of a narrow core on top of a broader diskline-like component.}
\vspace{0.2cm}
\label{fig_profile}
\end{figure}

The profile of the stacked iron residuals obtained with our analysis is shown in
Figure \ref{fig_profile}. It reveals a combination of a narrow core, seen almost
ubiquitously in local AGN (\citealt{Nandra07}), on top of a broader diskline-like
component. In order to analyse the stacked iron emission profile, we followed
\cite{Chaudhary12} and generated a powerlaw continuum with the same energy
binning as used above, then multiplied this continuum by the observed ratio
spectrum. We adopted a photon index of $\Gamma=1.8$ for the continuum, the
median of the distribution obtained from our powerlaw fits to the observed
sample. This was then imported into \xspec\ using \flx2xsp\ (part of the
\heasoft\ distribution).

\begin{figure}
\hspace*{-0.6cm}
\epsscale{1.175}
\plotone{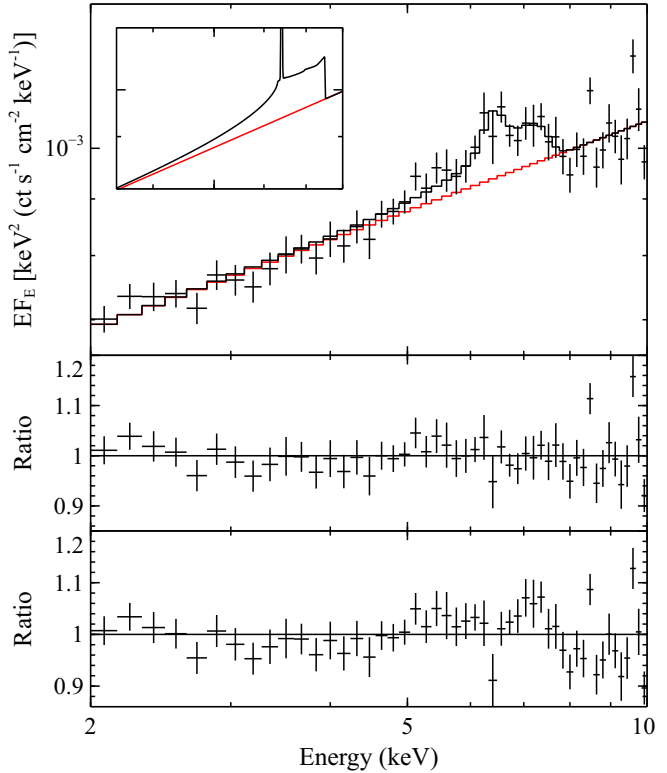}
\caption{
The model applied to the stacked 2--10\,\kev\ residuals after the
re-introduction of a $\Gamma=1.8$ powerlaw continuum (\textit{top}) and the
data/model residuals with this model applied (\textit{middle}). The total model is
shown in black, and the powerlaw continuum in red; the inset shows a zoom-in
on the model for the Fe K profile. The parameters adopted for the broad line
component are $a^*=0.7$, $i=45^{\circ}$, $q_{\rm{in}}=6$, $q_{\rm{out}}=3$,
$R_{\rm{br}}=6$\,\rg\ (see text), and we find a best-fit energy of
$E\sim6.7$\,\kev. The data/model residuals for a model including just a narrow
iron emission line are also shown (\textit{bottom}).
}
\vspace{0.2cm}
\label{fig_model}
\end{figure}

A simple powerlaw continuum provides a poor fit to the resulting spectrum
($\chi^{2}/\rm{DoF}=109/48$). We next include a narrow emission line from
neutral iron. However, given the range of redshifts, the rest-frame iron \ka\
lines occur at a range of observed energies. Differences in the spectral resolution
of the ACIS detectors at these energies will result in some apparent broadening
of narrow features beyond the instrumental resolution at 6\,\kev\ when stacking
spectra from different redshifts (\citealt{Iwasawa12}). Furthermore, our binning of
the individual spectra, along with our assumption that photons within these bins
are evenly distributed as a function of energy may also result in some broadening
of narrow features. In order to assess the combination of these effects we
performed a series of simulations (see section \ref{sec_sims}), and find that our
analysis procedure should cause an intrinsically narrow iron emission line to have
a width of $\sigma=0.165$\,\kev\ in the stacked spectrum. We therefore fix the
width of the narrow line to 0.165\,\kev\ in our analysis, and we also fix the
energy to that of neutral iron (6.4\,\kev). The addition of this line improves the fit
($\chi^{2}/\rm{DoF}=85/47$) and successfully accounts for the narrow core of
the observed emission profile. However, broad residuals still remain (see Figure
\ref{fig_model}).

We therefore include a second, broad Gaussian line, which further improves the
fit ($\chi^{2}/\rm{DoF}=61/44$). The broad component has 
$E=6.2\pm0.3$\,\kev\ and $\sigma=1.0^{+0.5}_{-0.3}$\,\kev\ (errors are
90\% confidence), significantly larger than the combined broadening of the
instrument and our analysis. We also construct a model in which the broad
Gaussian line is replaced by a relativistic emission line from the innermost
accretion disk, using the \relline\ model (\citealt{relconv}). Although we are
adopting a simplistic approach, modeling what would in reality be a combination
of direct continuum and a full reflection spectrum from the inner disk with a
powerlaw continuum and a single relativistic emission line, given the S/N more
detailed treatments are not warranted, but it is worth noting that the former will
also incorporate some contribution from the blurred reflection continuum that
would naturally accompany the line. For consistency, the \relline\ component is
smoothed with a $\sigma=0.165$\,\kev\ Gaussian, as per the broadening of
narrow features in our stacked spectrum. In this model, the details of the line
profile are determined by several parameters, and given the signal we are not
able to independently constrain them all. We therefore fix several of these
parameters to physically motivated values in our analysis, but we stress that the
results obtained do not strongly depend on our adopted parameter values.

The parameters we fix are the black hole spin, and the inclination and emissivity
profile of the accretion disk. We set the spin to be $a^{*}=0.7$, motivated by
both our previous results for high-redshift lensed quasars (\citealt{Reis14nat,
Reynolds14}), and the spin implied from the relation between BH spin and
radiative efficiency (\citealt{Thorne74}) and the average quasar radiative
efficiency of $\eta\sim0.1$ found by \cite{Soltan82}. We further assume that the
disk extends into the innermost stable circular orbit. The inclination is set to
45\deg, roughly similar to that expected for luminous, unobscured quasars.
Finally, we adopt a broken powerlaw profile for the radial emissivity of the disk,
with an inner index of $q_{\rm{in}}=6$, an outer index of $q_{\rm{out}}=3$, and
a break radius of $R_{\rm{br}}=6$\,\rg. This is motivated by both the compact
X-ray source sizes inferred for lensed quasars (\citealt{Dai10, Mosquera13,
Reis13corona, MacLeod15}) and the relativistic ray-tracing work of
\cite{Wilkins12}. The line energy and normalisation are free to vary, although the
energy is constrained to the range of iron \ka\ transitions (6.4--6.97\,\kev). The
inclusion of this \relline\ component formally provides the best fit of all the
models ($\chi^{2}/\rm{DoF}=60/45$), and the line energy obtained is
$E>6.6$\,\kev. The equivalent widths of the relativistic line and the
narrow core in our redshift-corrected stacked spectrum are
$EW_{\rm{rel}}=190\pm70$\,eV and $EW_{\rm{narrow}}=28^{+29}_{-27}$\,eV.
The strength of the relativistic emission is very similar to that expected for
reflection from a standard accretion disk (\citealt{George91}). In addition to
testing different spectral binning and re-sampling methods, as discussed above,
we have also systematically excluded each source and repeated the analysis on
the reduced sample, and find that these results are robust to the exclusion of any
individual source.

\section{Simulations}
\label{sec_sims}

In order to interpret these results, we also performed a series of simulations to
assess the detection significance of the broad component of the iron emission.
Using the \chandra\ responses, we simulated the spectra for \nsims\ samples of
27 sources using {\small{FAKEIT}} in \xspec, and performed the same stacking
analysis outlined above. We simulated an absorbed powerlaw continuum  with a
narrow ($\sigma=10$\,eV) iron line for each source, using the continuum
parameters obtained in our initial analysis of the real sample, an equivalent width
of $\sim$80\,eV and the accompanying \chandra\ exposure. No broad iron
emission component was included. We included counting statistics in our
simulations, and rebinned the individual simulated spectra to the same level as
the real data before performing the stacking analysis described above. For each
of the \nsims\ stacked ratio spectra obtained, we analyzed the simulated data in
the same manner as above, applying them to a $\Gamma=1.8$ powerlaw
continuum  and modeling the result in \xspec. We first model the stacked spectra
with a powerlaw continuum and a Gaussian emission line at 6.4\,\kev. The line
width is free to vary, in order to determine the combined broadening introduced
by our analysis procedure and the range of instrumental resolutions at the
observed Fe K energies (owing to the different source redshifts). As discussed
previously, we find the average width for the narrow line in the stacked spectra
to be $\sigma=0.165$\,\kev. We then add a \relline\ component with the
parameters used above, smoothed by a Gaussian with the width linked to that
found for the narrow line, and note the improvement in \chisq, in order to assess
the chance probability of obtaining the observed improvement in the instance
that no broad line is actually present. Of the \nsims\ samples simulated, only
\ngtr\ show a chance improvement equivalent to or greater than that observed,
implying that the detection significance of the broad feature detected in the real
data exceeds the 3$\sigma$ level.

\section{Discussion and Conclusions}
\label{sec_dis}

Following our recent detection of relativistic disk reflection features in two
lensed quasars beyond the local universe, RXJ1131-1231 ($z\sim0.658$;
\citealt{Reis14nat}) and Q2237+0305 ($z=1.695$; \citealt{Reynolds14}), we
have presented an analysis of a large sample of 27 lensed quasars in the
redshift range $1.0\lesssim{z}\lesssim4.5$ observed with \chandra, with a
combined exposure of over 1.6\,Ms. These do not currently have the archival
S/N of the two individual cases presented to date. While iron emission
has been detected from a few of the sources in our sample individually (\eg\
\citealt{Page04}), and there have previously been low-significance claims of
complexity in the iron bandpass for SDSS J0924+0219 and SDSS J1004+4112
(\citealt{Ota06, Chen12}), none have robust relativistic line detections, so
we focus on stacking the residuals in the iron K bandpass to simple powerlaw
AGN continuum models, similar to previous analyses (\citealt{Nandra97,
Guainazzi06, Chaudhary12}). We find that the stacked iron emission from this
sample shows both a narrow core and an underlying broad diskline-like
component, with a clear red wing to the emission profile (see Figure
\ref{fig_profile}). 

The presence of this broad component is robust to the various aspects of our
analysis, and is not dominated by the contribution from any one source. This is
an important point, as some of the sample may also individually exhibit
additional spectral complexities. For example, some \chandra\ observations of
PG1115+080 may show evidence for an X-ray outflow via high-energy iron
absorption lines at $\sim$7.4 and 8.5\,keV (rest-frame; \citealt{Chartas03}).
Five sources are lensed by either galaxy clusters or groups rather than individual
galaxies (\citealt{Chartas02, Grant04, Ota06, Ota12}), although in all these cases
the cluster/group emission is known to be negligible in the quasars iron
bandpass. Repeating the analysis excluding these five sources, we find including
\relline\ still gives a similar statistical improvement, but the data at the highest
energies has much lower S/N. The robustness of the detection of the broad
component to the exclusion of any individual source in turn implies it is also
robust to such details, and is genuinely representative of the average properties
of our sample.

We interpret this broad emission component as iron emission originating from
the inner accretion disks of these high-redshift sources. Although the S/N is
still low and other interpretations can almost certainly reproduce the data, for
example the right combination of complex absorption components could mimic
a diskline-like profile (\citealt{LMiller09}), our interpretation is motivated by
several recent results regarding local AGN. With the launch of the \nustar\
observatory (\citealt{NUSTAR}), high S/N broadband X-ray studies of local AGN
are now possible, and have demonstrated the presence of reflection from the
inner accretion disk in the Seyfert galaxy NGC1365 (\citealt{Risaliti13nat,
Walton14}). For the highest S/N local AGN, reverberation of the broad iron
emission has also now been detected (\citealt{Zoghbi12, Cackett14, Kara15}),
unambiguously demonstrating this emission arises through reprocessing of the
continuum close to the black hole. In addition, features consistent with
relativistic line emission are observed in both local active galaxies and Galactic
black hole binaries (\citealt{Walton12xrbAGN}), as expected if they arise from
the inner disk. Finally, one of the key advantages of lensed quasars is that the
size of the X-ray emitting region can be determined independently through
microlensing studies. These typically find the X-ray source to be compact
(\citealt{Chartas09, Dai10, Morgan12, Mosquera13}), broadly consistent with
the sizes inferred from disk reflection/reverberation from local sources.
We stress that while microlensing could potentially influence the precise details
of the iron emission profile, in particular the radial emissivity, it cannot
artificially broaden the line, as it only serves to magnify the intrinsic
emission. Regardless, by averaging both over time and over the sample, such
effects should naturally be reduced. 

Other stacking studies of moderate/high-redshift AGN in the iron bandpass
have also shown hints of broad emission (\citealt{Brusa05, Chaudhary12,
Iwasawa12, Falocco13}), although typically only at low significance. However,
these studies did not focus on lensed quasars, and were not able to take
advantage of the magnification of the intrinsic AGN emission.
Additionally, this lensing may preferentially enhance the iron emission
(\citealt{Chen12}), further aiding detection. Our analysis implies that relativistic
reflection from the inner disk may be relatively common in this lensed population.
This is a key point, as it demonstrates that with further observations this
population genuinely provides an opportunity to extend the sample of AGN spin
measurements out to high redshift with current instrumentation, beyond the two
cases presented to date (\citealt{Reis14nat, Reynolds14}). Such measurements
would allow us to begin providing observational tests of the basic coherent vs
chaotic growth scenarios, and ultimately of the more sophisticated models for
SMBH growth and galaxy formation (\citealt{Dubois14, Sesana14}).

\section*{ACKNOWLEDGEMENTS}

The authors would like to thank the reviewer for providing feedback which
helped improve this paper, and Julian Merten for useful discussions.
The work of DJW/DS was performed at JPL/Caltech, under contract with NASA.

{\it Facilites:} \facility{Chandra}



\bibliographystyle{../../mn2e}

\bibliography{../../references}

\label{lastpage}

\end{document}